\documentclass[twocolumn]{aastex631}

\newcommand{\objname}{2009 DQ$_{118}$}

\usepackage[outline]{contour}
\usepackage[percent]{overpic}
\newcommand{\labelcolor}{yellow}

\newcommand{\labelpicA}[5]{
\begin{overpic}[width=#4\linewidth]{#1}
	\put (5,7) {\huge\color{\labelcolor} \textbf{\contour{black}{#2}}}
	\put (45,8) {\large\color{\labelcolor} \textbf{\contour{black}{#3}}}
	\put (2,52) {\includegraphics[width=0.11\linewidth]{#5}}
\end{overpic}
}

\usepackage{lipsum}

\received{2023 July 31}
\revised{2023 September 18}
\accepted{2023 September 24}
\published{2023 October 25}
\submitjournal{ApJL}

\begin{document}

\title{Recurring Activity Discovered on Quasi-Hilda 2009 DQ118}

\correspondingauthor{William J. Oldroyd}
\email{woldroyd@nau.edu}

\author[0000-0001-5750-4953]{William J. Oldroyd}
\affiliation{Dept. of Astronomy \& Planetary Science, Northern Arizona University, PO Box 6010, Flagstaff, AZ 86011, USA}

\author[0000-0001-7335-1715]{Colin Orion Chandler}
\affiliation{Dept. of Astronomy \& the DiRAC Institute, University of Washington, 3910 15th Ave NE, Seattle, WA 98195, USA}
\affiliation{LSST Interdisciplinary Network for Collaboration and Computing, 933 N. Cherry Avenue, Tucson AZ 85721, USA}
\affiliation{Dept. of Astronomy \& Planetary Science, Northern Arizona University, PO Box 6010, Flagstaff, AZ 86011, USA}

\author[0000-0001-9859-0894]{Chadwick A. Trujillo}
\affiliation{Dept. of Astronomy \& Planetary Science, Northern Arizona University, PO Box 6010, Flagstaff, AZ 86011, USA}

\author[0000-0003-3145-8682]{Scott S. Sheppard}
\affiliation{Earth and Planets Laboratory, Carnegie Institution for Science, 5241 Broad Branch Road. NW, Washington, DC 20015, USA}

\author[0000-0001-7225-9271]{Henry H. Hsieh}
\affiliation{Planetary Science Institute, 1700 East Fort Lowell Rd., Suite 106, Tucson, AZ 85719, USA}
\affiliation{Institute of Astronomy and Astrophysics, Academia Sinica, P.O.\ Box 23-141, Taipei 10617, Taiwan}

\author[0000-0001-8531-038X]{Jay K. Kueny}
\affiliation{Dept. of Astronomy and Steward Observatory, University of Arizona, 933 North Cherry Avenue Rm. N204, Tucson, AZ 85721, USA}
\affiliation{Lowell Observatory, 1400 W Mars Hill Rd, Flagstaff, AZ 86001, USA}
\affiliation{Dept. of Astronomy \& Planetary Science, Northern Arizona University, PO Box 6010, Flagstaff, AZ 86011, USA}
\affiliation{National Science Foundation Graduate Research Fellow}

\author[0000-0002-6023-7291]{William A. Burris}
\affiliation{Dept. of Physics, San Diego State University, 5500 Campanile Drive, San Diego, CA 92182, USA}
\affiliation{Dept. of Astronomy \& Planetary Science, Northern Arizona University, PO Box 6010, Flagstaff, AZ 86011, USA}

\author[0000-0002-7489-5893]{Jarod A. DeSpain}
\affiliation{Dept. of Astronomy \& Planetary Science, Northern Arizona University, PO Box 6010, Flagstaff, AZ 86011, USA}

\author[0000-0003-2521-848X]{Kennedy A. Farrell}
\affiliation{Dept. of Astronomy \& Planetary Science, Northern Arizona University, PO Box 6010, Flagstaff, AZ 86011, USA}

\author[0000-0002-2204-6064]{Michele T. Mazzucato}
\affiliation{Royal Astronomical Society, Burlington House, Piccadilly, London, W1J 0BQ, UK}
\affiliation{Physical Sciences Group, Siena Academy of Sciences, Piazzetta Silvio Gigli 2, 53100 Siena, Italy}
\affiliation{Active Asteroids Citizen Scientist}

\author[0000-0002-9766-2400]{Milton K. D. Bosch}
\affiliation{Active Asteroids Citizen Scientist}

\author{Tiffany Shaw-Diaz}
\affiliation{Active Asteroids Citizen Scientist}

\author{Virgilio Gonano}
\affiliation{Active Asteroids Citizen Scientist}

\begin{abstract}

We have discovered two epochs of activity on quasi-Hilda \objname{}. Small bodies that display comet-like activity, such as active asteroids and active quasi-Hildas, are important for understanding the distribution of water and other volatiles throughout the solar system. Through our NASA Partner Citizen Science project, {\it Active Asteroids}, volunteers classified archival images of \objname{} as displaying comet-like activity. By performing an in-depth archival image search, we found over 20 images from UT 2016 March 8--9 with clear signs of a comet-like tail. We then carried out follow-up observations of \objname{} using the 3.5 m Astrophysical Research Consortium Telescope at Apache Point Observatory, Sunspot, New Mexico, USA and the 6.5 m Magellan Baade Telescope at Las Campanas Observatory, Chile. These images revealed a second epoch of activity associated with the UT 2023 April 22 perihelion passage of \objname{}. We performed photometric analysis of the tail and find that it had a similar apparent length and surface brightness during both epochs. We also explored the orbital history and future of \objname{} through dynamical simulations. These simulations show that \objname{} is currently a quasi-Hilda and that it frequently experiences close encounters with Jupiter. We find that \objname{} is currently on the boundary between asteroidal and cometary orbits. Additionally, it has likely been a Jupiter family comet or Centaur for much of the past 10 kyr and will be in these same regions for the majority of the next 10 kyr. Since both detected epochs of activity occurred near perihelion, the observed activity is consistent with sublimation of volatile ices. \objname{} is currently observable until $\sim$mid-October 2023. Further observations would help to characterize the observed activity.

\end{abstract}


\section{Introduction} \label{sec:intro}
The active asteroids are a population of small solar system bodies on asteroidal orbits, but which show signs of comet-like activity, such as tails \citep{jewittActiveAsteroids2015}. Fewer than 50 of these intriguing objects are known \citep{2022arXiv220301397J}, and their relative sparseness in the overall asteroid population ($>$10$^6$) remains unexplained.

The quasi-Hildas (also known as quasi-Hilda asteroids, quasi-Hilda objects, or quasi-Hilda comets, whether or not they display cometary activity) are related to the active asteroids. They orbit between the outer edge of the main asteroid belt and the Jupiter Family Comets (JFCs). Quasi-Hildas are also characterized as being near, but not within, the 3:2 interior mean-motion orbital resonance with Jupiter \citep{2006A&A...448.1191T}; the Hilda asteroid group is defined as being within this resonance. The quasi-Hildas have short dynamical lifetimes and some of them likely migrated to their current orbits from the outer solar system through interactions with the giant planets \citep{2016A&A...590A.111G}. Additionally, relatively few quasi-Hildas ($<$15) have been found to exhibit comet-like activity \citep{chandlerMigratoryOutburstingQuasiHilda2022} out of the $\sim$300 identified so far \citep{2016A&A...590A.111G}, with activity on many of these objects being discovered in recent years, for example, 2008 GO$_{98}$ \citep{2018P&SS..160...12G}, P/2010 H2 \citep{2020PSJ.....1...77J}, 282P \citep{chandlerMigratoryOutburstingQuasiHilda2022}, and 2018 CZ$_{16}$ \citep{2023RNAAS...7..106T}. Shoemaker-Levy 9, a comet known for its well-observed impact with Jupiter, was also likely a quasi-Hilda \citep{2008A&A...489.1355O}.

Comet-like activity on traditionally non-cometary bodies, such as asteroids, has revealed the presence of a previously unrecognized reservoir of volatile ices in our solar system \citep{2015Icar..248..289H}. The distribution of this material throughout the solar system is poorly understood and further study may shed light on pathways for delivery of these volatiles to Earth \citep{2000M&PS...35.1309M,2018SSRv..214...47O}. Additionally, volatile ices on small solar system bodies may provide crucial resource reservoirs for future space exploration \citep[see][and references therein]{chandlerChasingTailsActive2022}.

To study activity on asteroids and other bodies throughout the solar system, we created {\it Active Asteroids}\footnote{\url{activeasteroids.net}}, a NASA Partner Citizen Science project hosted on the {\it Zooniverse}\footnote{\url{zooniverse.org}} citizen science platform \citep{chandlerChasingTailsActive2022}. Activity was discovered on \objname{} as a result of this project \citep{2023RNAAS...7...42O}.

In this work we will summarize the initial detection of activity on \objname{} through the {\it Active Asteroids} project followed by a description of our archival search for additional images containing activity.
Next, we discuss our follow-up observations and photometric analysis of \objname{}, as well as the discovery of a second epoch of cometary activity.
We also present a dynamical analysis of the orbital evolution of \objname{} and compare it with other known active quasi-Hildas. Finally, we discuss mechanisms that could cause activity on \objname{}, as well as future observing opportunities regarding this object.

\section{Citizen Science Discovery} \label{sec:citsci}
In order to better study the active asteroids, we seek to discover more of them through our {\it Active Asteroids} Citizen Science project. For this project, we retrieve publicly-available images of known asteroids and other small solar system bodies from the Dark Energy Camera (DECam) archive. The wide field of view of DECam ($2.2^{\circ} \times 2.2^{\circ}$) on the 4 m Blanco telescope \citep{2008SPIE.7014E..0ED} is excellently situated for detecting activity. After employing our automated vetting process described in \citet{chandlerSAFARISearchingAsteroids2018} and \citet{chandlerChasingTailsActive2022}, images passing our data quality metrics are examined by {\it Active Asteroids} volunteers and classified by them as either active or inactive. 

Images identified as containing activity by citizen scientists are then reviewed by our science team to further validate candidate detections. Next, we perform an in-depth archival search on promising candidates from this activity identification process \citep{chandlerChasingTailsActive2022}. These searches yield additional images displaying activity for some candidate objects, allowing us to further study potential mechanisms for activity.

\begin{figure*}[htp]
    \centering
    \begin{tabular}{cccc}
        \labelpicA{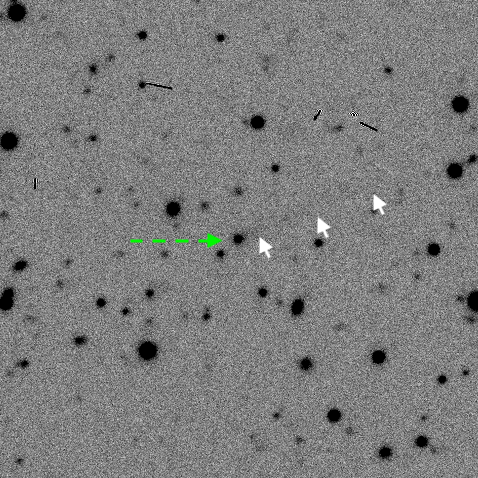}{}{}{0.23}{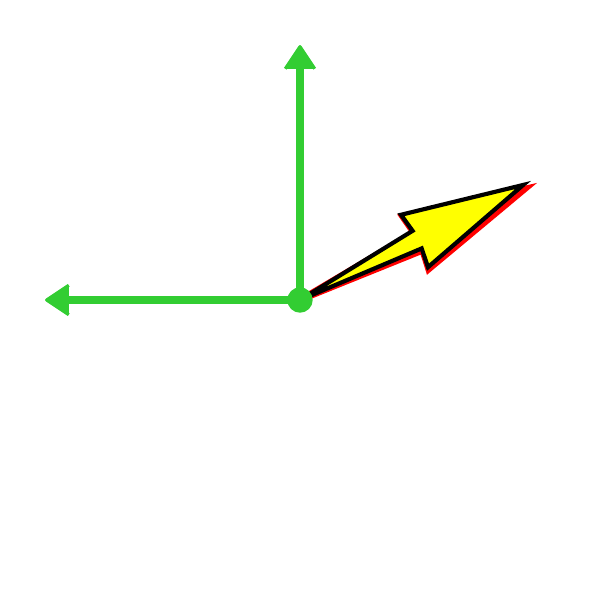} &
        \labelpicA{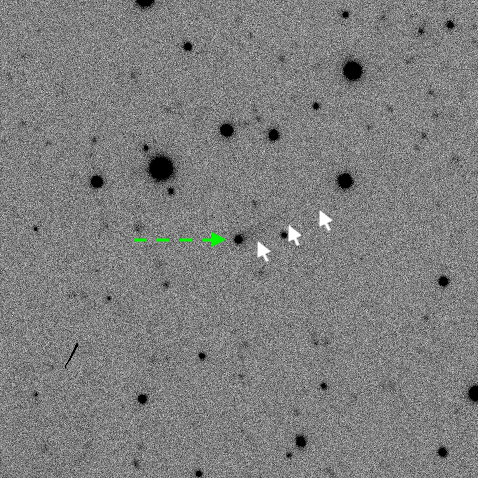}{}{}{0.23}{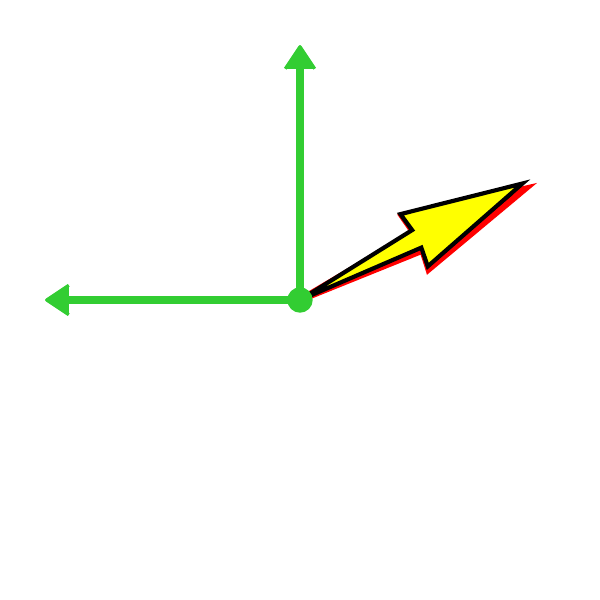} &
        \labelpicA{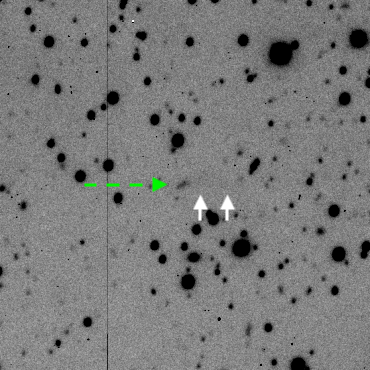}{}{}{0.23}{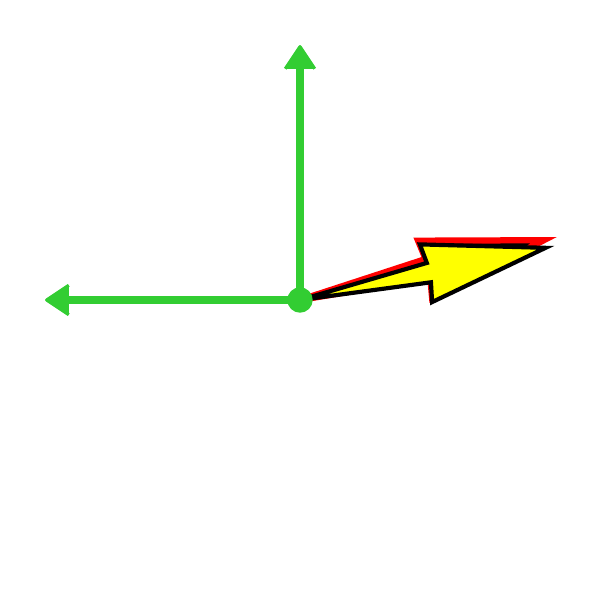} &
        \labelpicA{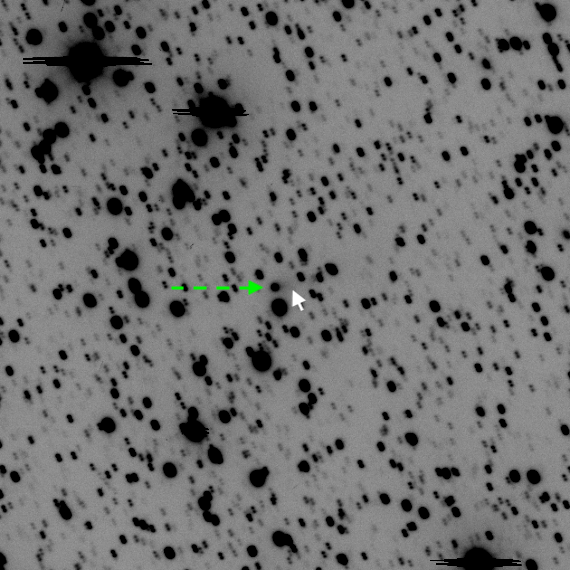}{}{}{0.23}{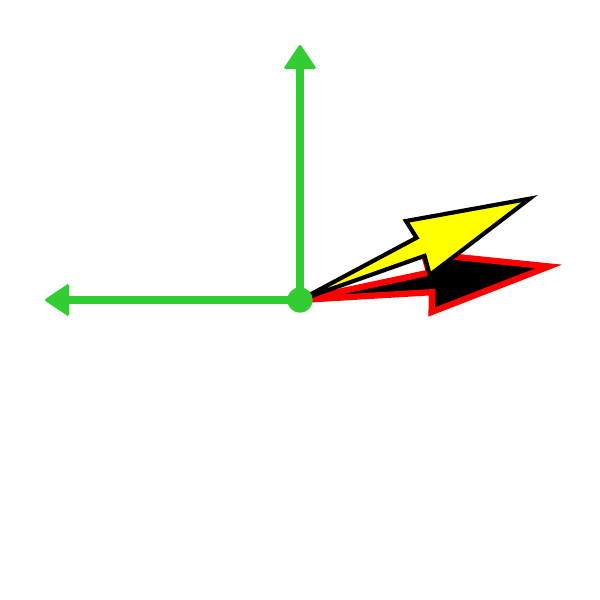} \\
        (a) & (b) & (c) & (d)\\
    \end{tabular}
    \caption{Images of \objname{} (green dashed arrows) displaying a cometary tail (white arrows). Frames (a) and (b) are from the first activity epoch and resulted from our {\it Active Asteroids} citizen scientist project and archival search. Frame (c) is an Apache Point Observatory (APO) follow-up image showing faint signs of activity resulting in the tentative discovery of the second epoch of activity. In frames (a) through (c), the negative heliocentric velocity (black arrow outlined in red) and anti-solar (yellow arrow) directions projected to the on-sky plane coincide with each other and the direction of the tail. Frame (d) is a stack of our Magellan follow-up observations confirming the discovery of the second activity epoch. In this frame, the tail is oriented between the anti-solar (yellow arrow) and negative heliocentric velocity (black arrow outlined in red) directions projected to the on-sky plane. North is up and East is left in each image (solid green arrows) and all directions are referenced to the ephemeris location of \objname{} (which is centered in each image) at the time of observation as given by JPL Horizons \citep{giorginiJPLOnLineSolar1996}. 
    {\bf (a):} 300 s {\it VR}-band Dark Energy Camera (DECam) image taken with the 4 m Blanco telescope at Cerro Tololo Inter-American Observatory (CTIO), Chile on UT 2016 March 8 (Prop. ID 2016A-0189; PI Rest; observers A. Rest, DJJ).
    {\bf (b):} 200 s {\it r}-band DECam image, UT 2016 March 9 (Prop. ID 2015A-0121; PI von der Linden; observer A. von der Linden).
    {\bf (c):} 300 s {\it VR}-band image taken with the Astrophysical Research Consortium Telescope Imaging Camera (ARCTIC) on the APO 3.5 m Astrophysical Research Consortium (ARC) Telescope, UT 2023 February 24 (Prop. ID 2Q2023-UW08; PI Chandler; observer C. Chandler).
    {\bf (d):} A co-added stack of four 150s {\it WB4800-7800}-band images taken with the Inamori-Magellan Areal Camera \& Spectrograph (IMACS) on the 6.5 m Magellan Baade Telescope at Las Campanas Observatory, Chile on UT 2023 April 22 (PI S. Sheppard; observer S. Sheppard).
    }
\label{fig:images}
\end{figure*}

As a result of the {\it Active Asteroids} project, we discovered comet-like activity originating from \objname{} \citep[as reported in our preliminary announcement][]{2023RNAAS...7...42O}. Once volunteers had identified \objname{} as active, our archival search produced over 20 images of \objname{} displaying a tail. All of these images showing activity were from UT 2016 March 8--11, just 4 months before its 2016 perihelion passage (heliocentric distance $r_h = 2.55$ au, true anomaly $f = 322^{\circ}$). We also identified $\sim$10 images without readily apparent signs of activity. All of these inactive images were taken more than a year away from \objname{} perihelion passages. Representative active images from this search are shown in Figure~\ref{fig:images}.

\section{Follow-up Observations} \label{sec:obs}
We acquired follow-up observations of \objname{} using the Astrophysical Research Consortium Telescope Imaging Camera (ARCTIC) on the Apache Point Observatory (APO) 3.5 m Astrophysical Research Consortium (ARC) telescope in Sunspot, New Mexico, USA \citep{2016SPIE.9908E..5HH}. On UT 2023 February 24 we took 12 300 s {\it VR}-band images of \objname{} (Prop. ID 2Q2023-UW08; PI Chandler; observer C. Chandler). The conditions were poor, with intermittent clouds and a seeing of $\sim$2.7$\arcsec$. On this date, \objname{} was approaching perihelion ($r_h = 2.456$ au, $f = 343.0^{\circ}$) and we saw faint indications of a tail (Figure \ref{fig:images}(c)). 

In order to confirm this second epoch of activity, we acquired follow-up observations of \objname{} using the Inamori-Magellan Areal Camera \& Spectrograph (IMACS) on the 6.5 m Magellan Baade telescope at Las Campanas Observatory, Chile \citep{2011PASP..123..288D}. Our observations, taken on UT 2023 April 22 (PI: S. Sheppard), were comprised of four 150 s images in the broad {\it WB4800-7800} filter (similar to a {\it VR} filter) with seeing between 0.8$\arcsec$ and 0.9$\arcsec$ and a pixel scale of 0.2$\arcsec$/pixel. These observations were timed so that \objname{} was at perihelion ($r_h = 2.430$ au, $f = 359.9^{\circ}$) since this is an ideal time to check for signs of cometary activity. 

Our observations show a faint tail originating from \objname{} and oriented in the direction of the anti-solar and negative heliocentric velocity vectors projected to the on-sky plane (Figure \ref{fig:images}).
We performed photometry on images from both epochs in order to compare the tail between the apparitions. Epoch 1 DECam data were calibrated using Pan-STARRs DR1 {\it r}-band data, whereas epoch 2 Magellan data were calibrated to Gaia EDR3 {\it G}-band measurements ({\it r}-band catalogs were unavailable for this location) and transformed to the equivalent Sloan {\it r}-band \citep[a calibration proxy used for comparing IMACS {\it WB4800-7800} data with other data sets, see][]{2022PSJ.....3..175P} using the $\mathrm{{\it G}}_{\mathrm{{\it BP}}}-\mathrm{{\it G}}_{\mathrm{{\it RP}}}$ colors as described in the Gaia Early Data Release 3 Documentation\footnote{\url{https://gea.esac.esa.int/archive/documentation/GEDR3/Data_processing/chap_cu5pho/cu5pho_sec_photSystem/cu5pho_ssec_photRelations.html}, Table 5.7.}. We then compared the resulting magnitudes with the expected extinction corrected magnitudes reported by JPL Horizons \citep{giorginiJPLOnLineSolar1996}, thus accounting for phase correction, transforming from {\it V}-band to {\it r}-band assuming solar colors as in \citet{2019AJ....157...54J}, using the transformations given by \citet{2006A&A...460..339J}. In epoch 1, \objname{} had an {\it r}-band magnitude of 20.7, and it had an equivalent {\it r}-band magnitude of $\sim$20.3 during epoch 2; 0.4 magnitudes brighter than expected in both epochs. The tail was roughly 18$\arcsec$ long during epoch 1 and it had a surface brightness of 24.4 mag/arcsec$^2$. During epoch 2, \objname{} was in a crowded field which complicated measurement of the tail. We place a lower limit of 9$\arcsec$ on the length of the tail in epoch 2 with a maximum length of approximately 21$\arcsec$, and a surface brightness of 24.3 mag/arcsec$^2$. Hence, the tail had a similar apparent length and surface brightness during both epochs.

The detection of two separate epochs of activity on \objname{} likely points to sublimation of volatile ices as the primary activation mechanism. Because of the proximity of \objname{} to its perihelion passage on UT 2023 April 22,  observations during this observing window will be particularly useful for further characterization of the tail if they reach a depth of $V\gtrsim$ 23 (sufficiently deep to detect the tail). \objname{} will be observable through $\sim$mid October 2023 (especially from the southern hemisphere), and then again in mid 2024, albeit, not near perihelion ($f\approx 100^{\circ}$).

\section{Dynamical Analysis} \label{sec:dynamics}
In order to further study potential causes for activity on \objname{}, we performed a set of dynamical simulations to examine its short term (1-10 kyr) orbital history and future. For these simulations we utilized the {\tt IAS15} integrator \citep{2015MNRAS.446.1424R} from the {\tt REBOUND} $N$-body integration package \citep{2012A&A...537A.128R} in  {\tt Python}. To account for observational uncertainties in its orbit, we created 500 dynamical clones of \objname{}. These clones were drawn from Gaussian distributions using the orbital elements and uncertainties (see Table \ref{tab:elements}) from the JPL Horizons Small Body Database \citep{giorginiJPLOnLineSolar1996}. We integrated each orbital clone for $\pm$10 kyr along with the Sun and planets \citep[except Mercury, which has a negligible impact and requires much more computation time to properly simulate, see, for example,][]{2023MNRAS.tmp..703B,2022MNRAS.510.4302H} with a timestep of 0.02 yr, sufficient for resolving the orbit of Venus \citep[see][for discussion on adequate orbital resolution]{2022MNRAS.510.4302H,2015AJ....150..127W}.

\begin{deluxetable}{lrcc}[!h]
    \tablenum{1}
    \tablecaption{\objname{} Orbital Parameters}
    \label{tab:elements}
    \tablewidth{0pt}
    \tablehead{
    \colhead{Parameter} & \colhead{Value} & \colhead{Uncertainty} & \colhead{Units}
    }
    \startdata
    Semi-major axis $a$ & 3.577 & 1.608$\times10^{-7}$ & au \\
    Eccentricity $e$ & 0.321 & 1.226$\times10^{-7}$ & - \\
    Inclination $i$ & 9.391 & 2.101$\times10^{-5}$ & deg \\
    Longitude of the & 344.658 & 7.191$\times10^{-5}$ & deg \\
    ascending node $\Omega$ & & & \\
    Argument of & 252.202 & 9.071$\times10^{-5}$ & deg \\
    perihelion $\omega$ & & & \\
    Mean anomaly & 351.749 & 5.298$\times10^{-5}$ & deg \\
    at epoch $M$ & & & \\
    Perihelion distance $q$ & 2.430 & 4.692$\times10^{-7}$ & au \\
    Aphelion distance $Q$ & 4.723 & 2.124$\times10^{-7}$ & au \\
    Orbital period $P$ & 6.765 & 4.562$\times10^{-7}$ & yr \\
    Tisserand parameter & 3.004 & 1.199$\times10^{-7}$ & - \\
    w.r.t.~Jupiter $T_\mathrm{J}$ & & & \\
    \enddata
    \vspace{8pt}
    {\bf Note:} Data acquired on UT 2023 June 14 from the JPL Horizons Small Body Database \citep{giorginiJPLOnLineSolar1996}. Epoch TDB 2023 February 25. JPL solution date PST 2022 March 8. Tisserand parameter calculated using Equation \ref{eq:TJ}. Uncertainties reported are 1$\sigma$.\\ 
    \vspace{8pt}
\end{deluxetable}

\begin{figure*}[htp]
    \centering
    \begin{tabular}{cc}
        (a) Plan view & (b) Log Distance from Jupiter\\
        \includegraphics[width=0.43\linewidth]{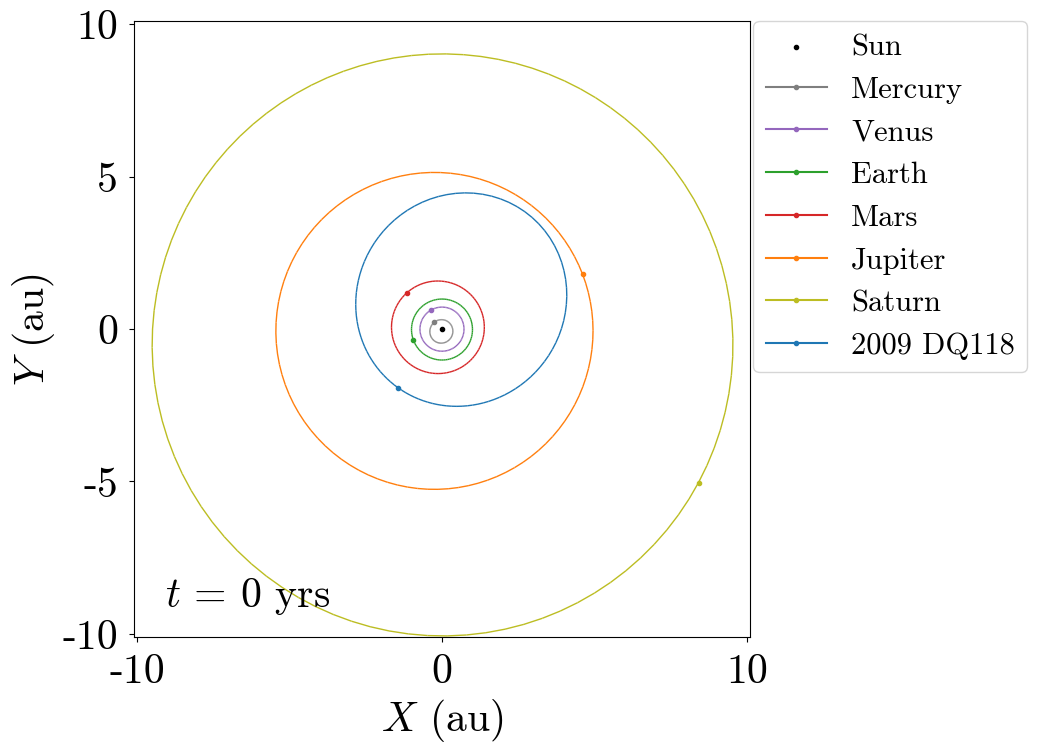} & \includegraphics[width=0.45\linewidth]{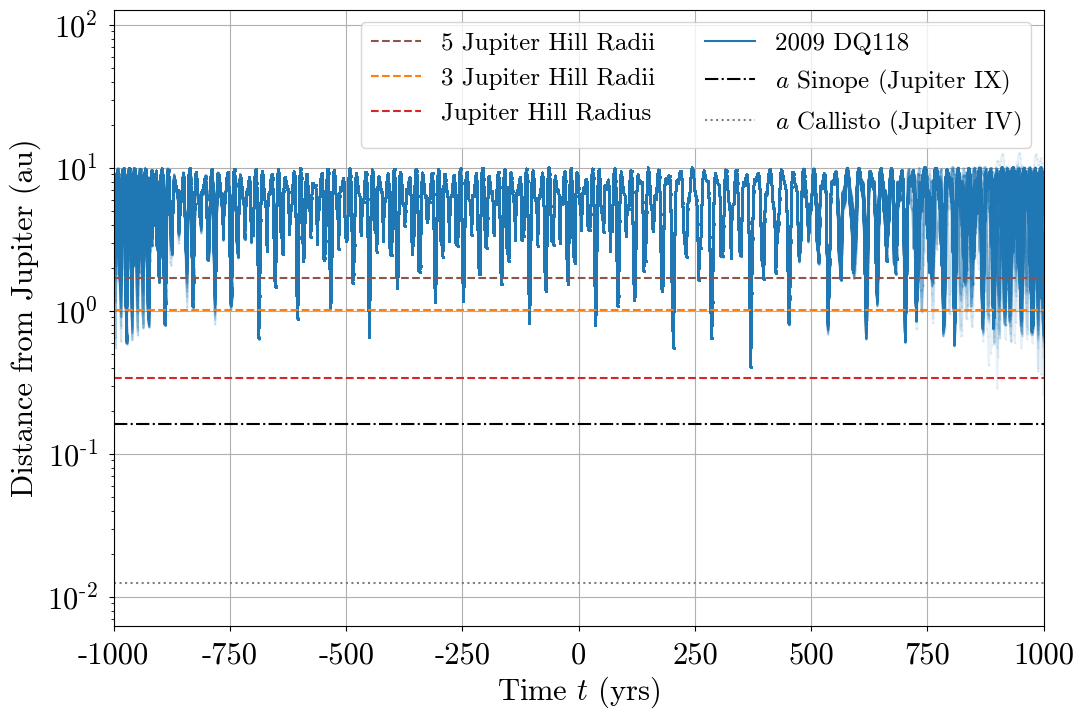} \\
        (c) Heliocentric Distance & (d) Dynamical Class\\
        \includegraphics[width=0.45\linewidth]{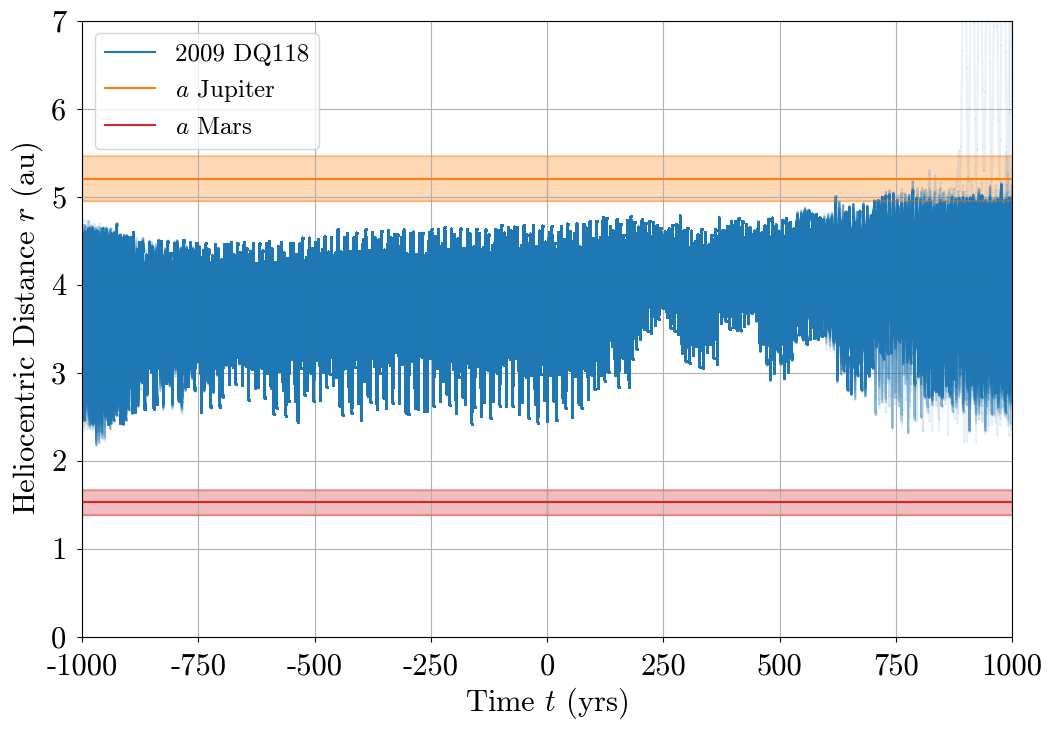} & \includegraphics[width=0.45\linewidth]{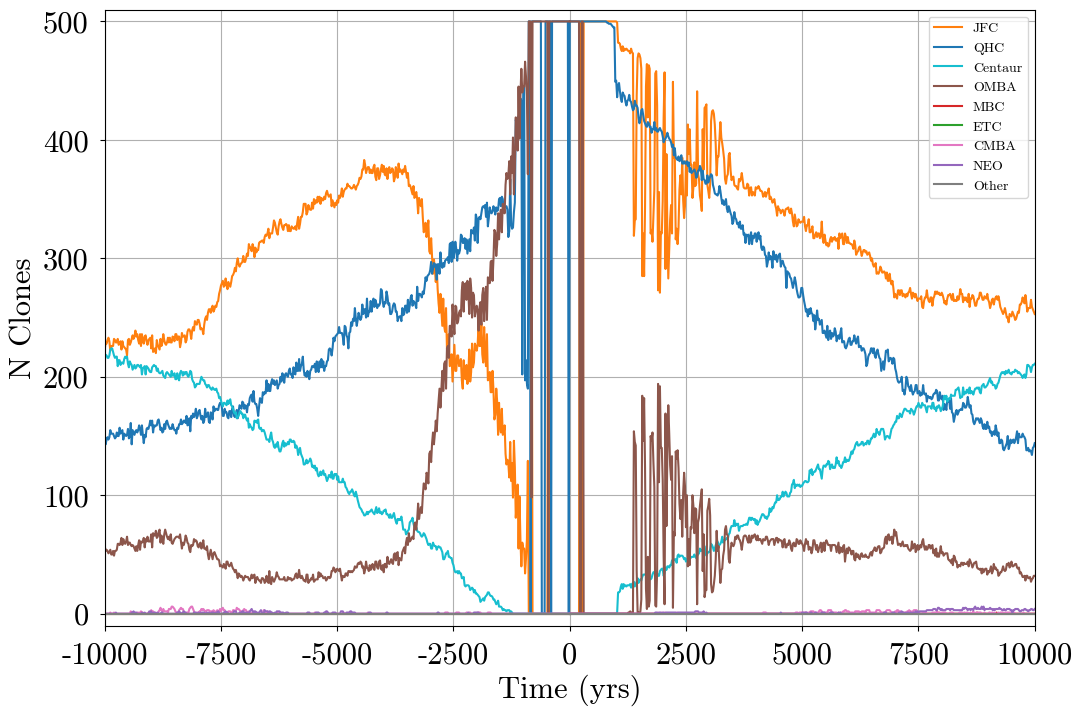} \\
    \end{tabular}
    \caption{Dynamical evolution of \objname{} orbital clones indicating changes to its orbit and dynamical class over short timescales.
    {\bf (a):} Orbits of \objname{} and the planets at $t=0$, UT 2023 April 11. Note the proximity of the orbits of \objname{} and Jupiter. 
    {\bf (b):} Log distance between \objname{} orbital clones and Jupiter as a function of time. Distances of 5, 3, and 1 Hill radii are marked to emphasize increasingly strong perturbations from close encounters. Semi-major axes of two Jovian moons are given for reference. Note that strong downward spikes, representing close encounters, correspond with changes in the orbit. Also note the onset of dynamical chaos before $\sim-750$ yr and after $\sim600$ yr. 
    {\bf (c):} Heliocentric distance of \objname{} orbital clones. Note how \objname{} clones begin to cross within the perihelion distance of Jupiter (shaded orange region below the orange line) after $\sim600$ yr.
    {\bf (d):} Dynamical class of \objname{} over $\pm$10 kyr (vs $\pm$1 kyr for panels (b) and (c)). These values correspond to those given in Table \ref{tab:outcomes}.
    }
    \label{fig:dynamics}
\end{figure*}

As a result of our simulations, we find that \objname{} experiences frequent close encounters with Jupiter over a $\pm$1,000 yr timescale. Many of these encounters are within 2-3 Hill radii of Jupiter (Figure \ref{fig:dynamics}), where the Hill radius \citep{hillResearchesLunarTheory1878} is computed as 
\begin{equation}
    r_\mathrm{H} \approx a(1-e)(m/3M)^{1/3},
    \label{282P:eq:rH}
\end{equation}
where $a$, $e$, and $m$ are the semi-major axis, eccentricity, and mass of the secondary body respectively (Jupiter in this case), and $M$ is the mass of the primary body (the Sun). For Jupiter, the Hill radius is $r_{\mathrm{H, J}} \approx 0.34$ au. An object passing within a few Hill radii of a planet will be subject to strong gravitational perturbations that will likely alter the orbit of the small body. This is the case for \objname{}, which has had recent changes in its orbit due to these encounters and will continue to have orbit-changing encounters in the near future as shown in Figure \ref{fig:dynamics}.

The proximity of the orbits of \objname{} and Jupiter is also connected to the Tisserand parameter with respect to Jupiter $T_\mathrm{J}$ of \objname{}. The Tisserand parameter with respect to Jupiter is a mostly-constant metric for the strength of the gravitational effect of Jupiter on the orbit of another body. It is defined as
\begin{equation}
	T_\mathrm{J} = \frac{a_\mathrm{J}}{a} + 2\cos(i)\sqrt{\frac{a}{a_\mathrm{J}}\left(1-e^2\right)},
	\label{eq:TJ}
\end{equation}
where $a$, $e$, and $i$ are the semi-major axis, eccentricity, and inclination, respectively, of the small body and $a_\mathrm{J}$ is the semi-major axis of Jupiter. Small bodies are often categorized based on $T_\mathrm{J}$, with objects that have $T_\mathrm{J}>3$ being classified as asteroids (which do not cross the orbit of Jupiter) while those with $T_\mathrm{J}<3$ are considered comets \citep[Jupiter orbit crossing;][]{1996ASPC..107..173L}. 
\objname{} has a $T_\mathrm{J}$ of 3.004, right on the traditional $T_\mathrm{J}=3$ boundary between asteroidal and cometary orbits. Additionally, although $T_\mathrm{J}$ is typically thought of as constant for a given object \citep{1972IAUS...45..503K}, close encounters with Jupiter cause minor changes to the $T_\mathrm{J}$ of \objname{}. These small changes cause \objname{} to cross $T_\mathrm{J}=3$ dozens of times over the course of $\pm$1,000 yr. However, these $T_\mathrm{J}$ crossings do not represent a dramatic orbital shift from one dynamical class to another, but rather serve to muddle the classification of \objname{} as seen by the abrupt jumps near $t=0$ in Figure \ref{fig:dynamics} (d).

\begin{deluxetable}{lccccc}[!h]
    \tablenum{2}
    \tablecaption{\objname{} Orbit Classification}
    \label{tab:outcomes}
    \tablewidth{0pt}
    \tablehead{
    \colhead{Orbit Class} & \colhead{$-10$ kyr} & \colhead{$-1$ kyr} & \colhead{$t$=0} & \colhead{1 kyr} & \colhead{10 kyr}
    }
    \startdata
    Centaur & 44\% & 0\% & 0\% & 0\% & 42.2\% \\
    JFC & 44.6\% & 9.2\% & 0\% & 100\% & 50.6\% \\
    Asteroid & 11.4\% & 90.8\% & 100\% & 0\% & 6.4\% \\
    NEO & 0\% & 0\% & 0\% & 0\% & 0.8\% \\
    \hline
    QH & 30.4\% & 87.6\% & 100\% & 88.2\% & 28.8\% \\
    \enddata
    \vspace{8pt}
    {\bf Note:} NEO stands for Near-Earth Object and QH for Quasi-Hilda. Percentages are calculated based on the number of orbital clones within the corresponding orbital class at the given times. Because quasi-Hilda is a non-exclusive pseudo-class, objects can be classified as either a quasi-Hilda and an asteroid, a quasi-Hilda and a JFC, or not a quasi-Hilda, but potentially still a JFC or asteroid.\\
    \vspace{8pt}
\end{deluxetable}

The results shown in Table \ref{tab:outcomes} highlight that although \objname{} is currently on an asteroidal orbit ($T_\mathrm{J}>3$), it is likely that it was either a JFC or even a Centaur \citep[$a_\mathrm{Jupiter} < q < a_\mathrm{Neptune}$; as in, for example,][]{2003AJ....126.3122T} for much of the past 10 kyr. Additionally, \objname{} will most likely be on a JFC or Centaur orbit for much of the coming 10 kyr (Figure \ref{fig:dynamics} (d)). There is a non-negligible probability, however, that \objname{} has been an asteroid for over 10 kyr and that, after becoming a JFC in the next 1,000 yr, it may transition back onto an asteroidal orbit or even become a near-Earth object.

\section{Discussion} \label{sec:disc}
\subsection{Dynamical Classification}
\begin{figure*}[htp]
    \centering
    \begin{tabular}{ccc}
        (a) Asteroid & (b) Near-Earth Asteroid & (c) Centaur \\
        \includegraphics[width=0.3\linewidth]{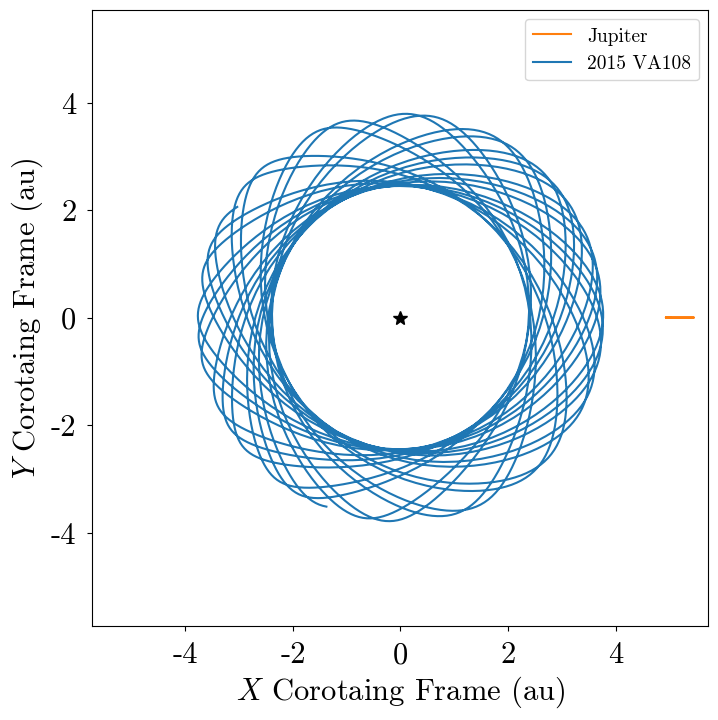} & \includegraphics[width=0.3\linewidth]{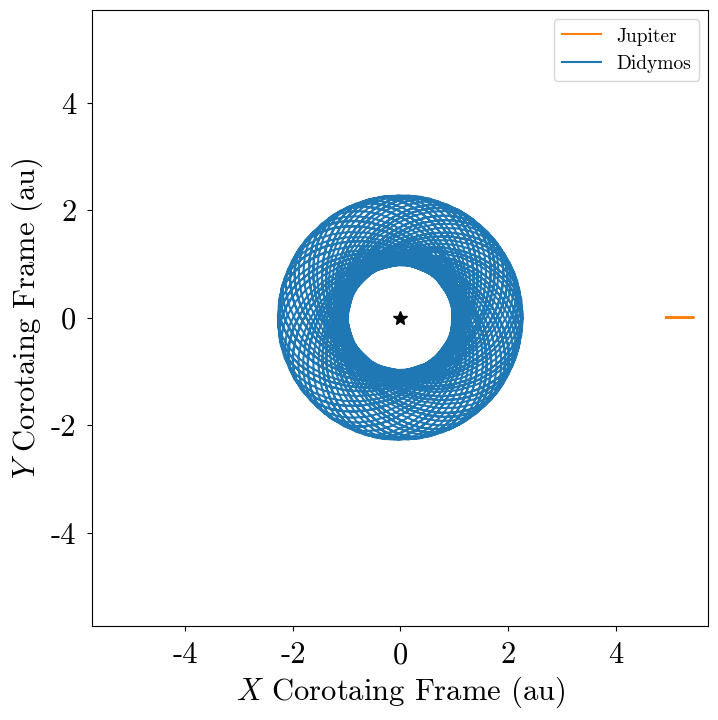} & \includegraphics[width=0.3\linewidth]{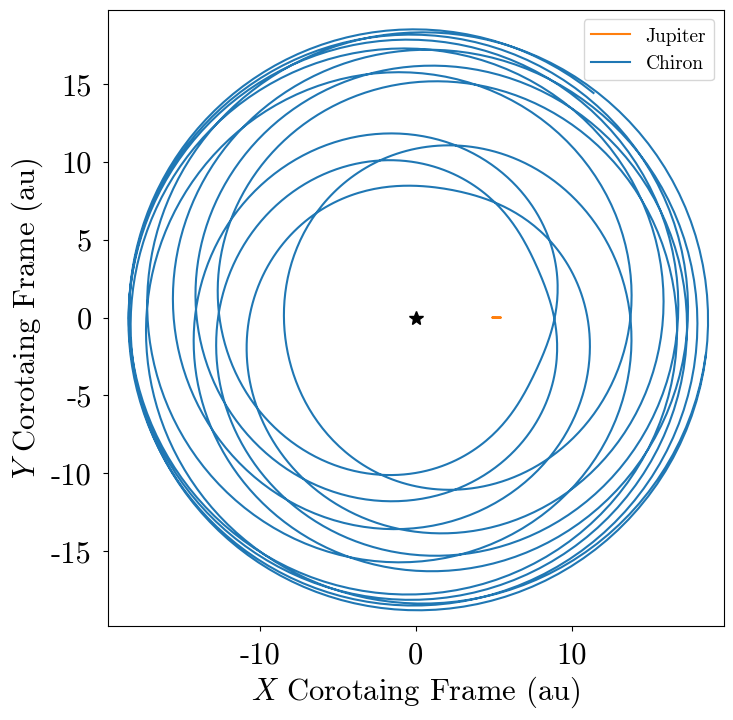} \\
        (d) Jupiter Family Comet & (e) Long Period Comet & (f) Trojan Asteroid \\
        \includegraphics[width=0.3\linewidth]{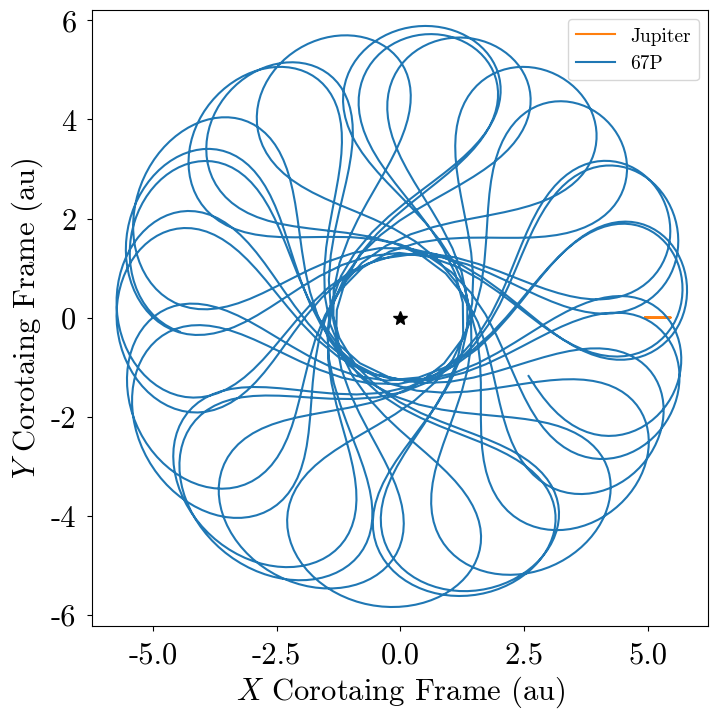} & \includegraphics[width=0.3\linewidth]{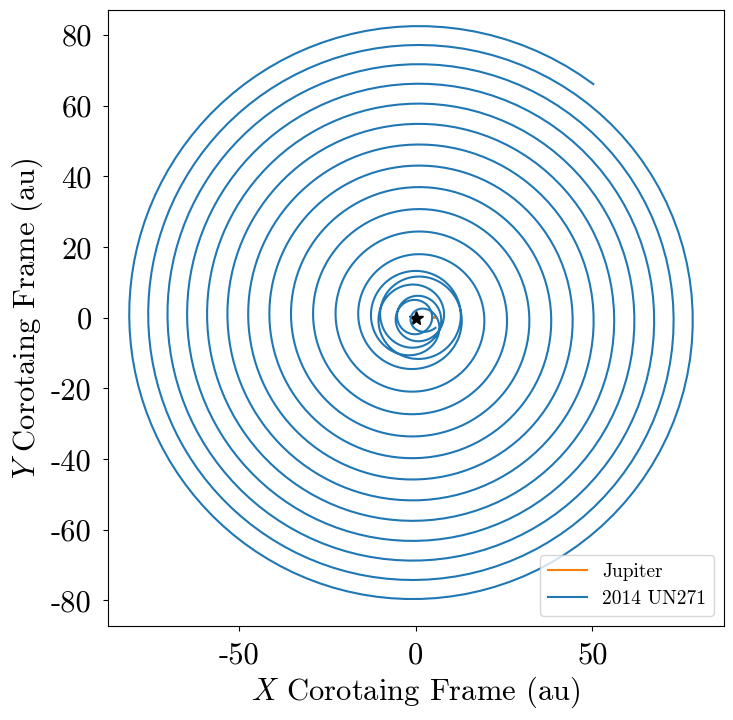} & \includegraphics[width=0.3\linewidth]{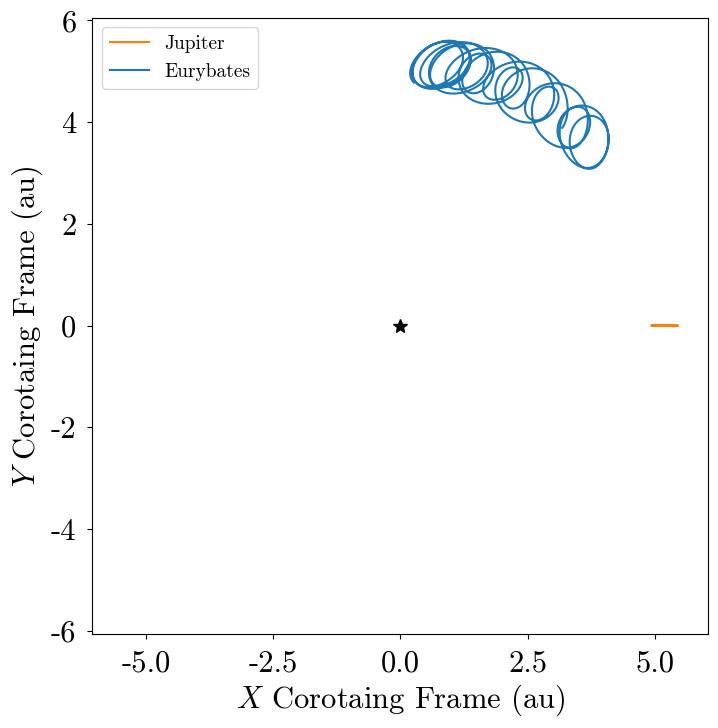} \\
        (g) Hilda Asteroid & (h) Quasi-Hilda & (i) Quasi-Hilda \\
        \includegraphics[width=0.3\linewidth]{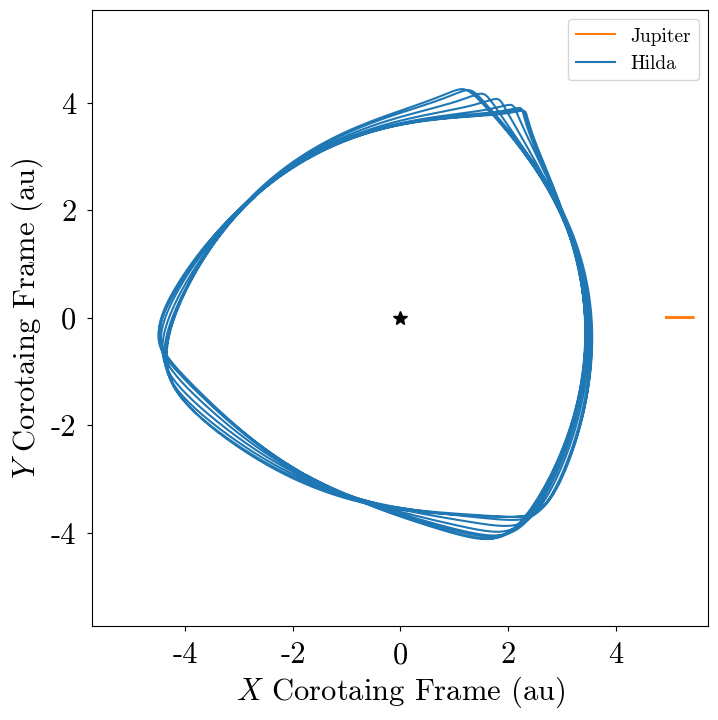} & \includegraphics[width=0.3\linewidth]{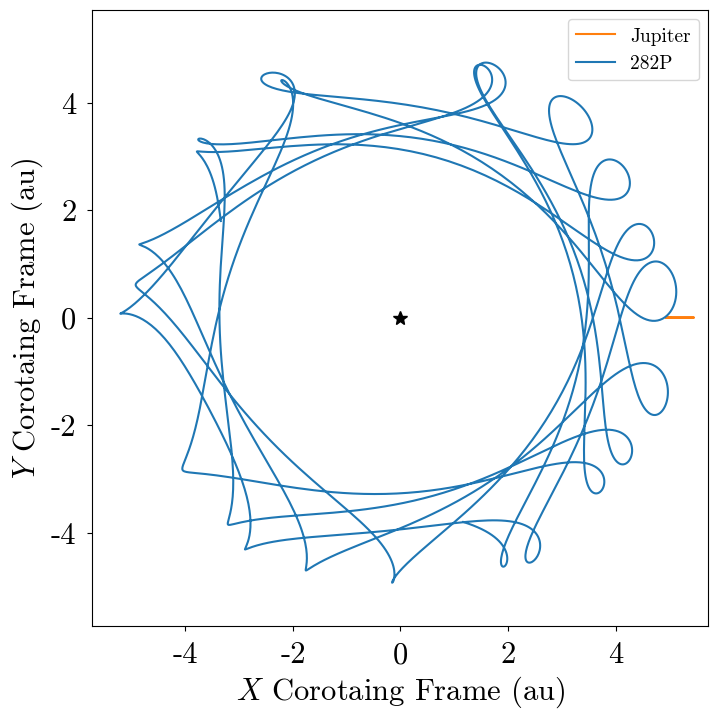} & \includegraphics[width=0.3\linewidth]{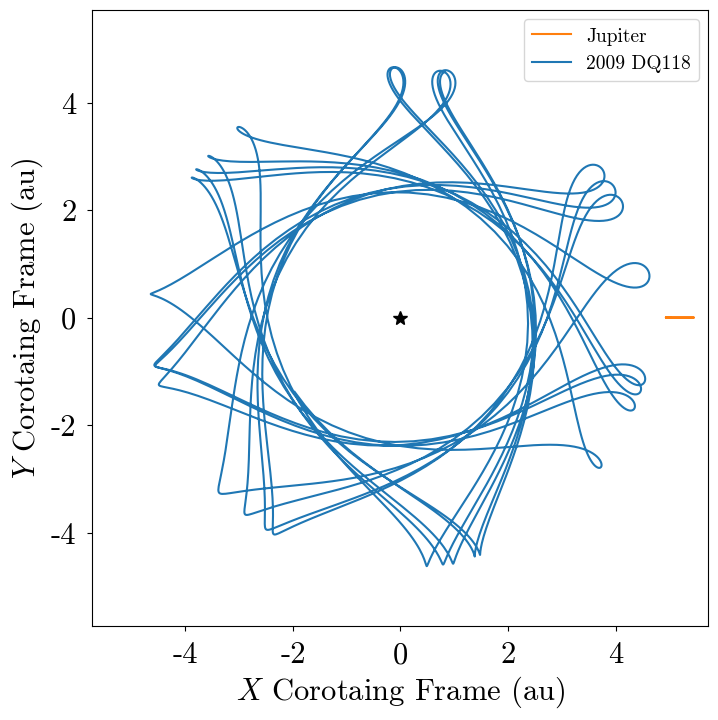} \\
    \end{tabular}
    \caption{Orbits of representative bodies (blue curves) from eight dynamical classes in the corotating frame with Jupiter (orange line) illustrating the similarities between \objname{} and other quasi-Hildas. Each subplot shows 200 yr of orbital integration in this reference frame.
    {\bf (a):} Active asteroid 2015 VA$_{108}$ orbits in the main asteroid belt and is a candidate main-belt comet \citep{2023RNAAS...7...27C}.
    {\bf (b):} Near-Earth binary asteroid (65803) Didymos-Dimorphos was the target of the NASA Double-Asteroid Redirection Test (DART) mission. It is the first artificial active asteroid \citep{DART-AA}.
    {\bf (c):} Active Centaur (2060) Chiron (95P) resides between the orbits of Jupiter and Uranus.
    {\bf (d):} Jupiter family comet 67P/Churyumov–Gerasimenko crosses the orbits of Jupiter and Mars. It was visited by the ESA Rosetta spacecraft.
    {\bf (e):} Long period comet C/2014 UN$_{271}$ (Bernardinelli–Bernstein) is currently inbound from the Oort cloud and will reach its perihelion, near the orbit of Saturn, in January 2031. Because this comet is highly inclined ($i\approx95^{\circ}$), it appears to be interior to the orbit of Jupiter in part of this X-Y projection.
    {\bf (f):} Trojan asteroid (3548) Eurybates in a characteristic Trojan tadpole orbit indicative of a 1:1 mean-motion resonance with Jupiter. Eurybates is a target of the NASA Lucy spacecraft mission.
    {\bf (g):} Asteroid (153) Hilda in its iconic 3:2 interior mean-motion resonance with Jupiter. Hilda asteroids are defined as being in this resonance and also display this trilobate pattern in this frame.
    {\bf (h):} Active quasi-Hilda  282P/(323137) 2003 BM80 displays a typical asymmetric quasi-Hilda corotating pattern \citep{chandlerMigratoryOutburstingQuasiHilda2022}.
    {\bf (i):} \objname{} with a quasi-Hilda orbit similar to 282P.
    }
    \label{fig:corotation}
\end{figure*}

In addition to its residence on the tenuous $T_\mathrm{J}=3$ asteroid-comet boundary, \objname{} exhibits many dynamical similarities to other active quasi-Hildas, such as 282P \citep{chandlerMigratoryOutburstingQuasiHilda2022}.
At $a \approx$ 3.6 au, \objname{} sits slightly outside of the quasi-Hilda semi-major axis range of $\sim$3.7 au $<a<$ $\sim$4.2 au given by \citet{2006A&A...448.1191T}, placing it closer to the Cybele asteroid group than to the Hildas. However, \objname{} experiences short-term dynamical evolution we find to be characteristic of the quasi-Hilda population. One simple way of visualizing the similarities between quasi-Hildas, in contrast with objects of other nearby dynamical classes, is by examining the orbits of these objects in the corotating frame with Jupiter. In Figure \ref{fig:corotation}, we show the orbits of objects in a frame that rotates at the same rate as the orbital motion of Jupiter so that Jupiter remains on the x-axis. Objects in separate dynamical classes appear quite different from one another in this frame, while objects in the same dynamical class have obvious similarities. Since \objname{} clearly resembles other quasi-Hildas when examined in the frame corotating with Jupiter, we classify \objname{} as a quasi-Hilda.

\subsection{Activity Mechanisms}
Among the various mechanisms for causing cometary activity on a small solar system body, the most well-studied is sublimation of volatile ices. Sublimation is the primary driver of activity on comets throughout the solar system. It is also used as a primary method for distinguishing between main-belt comets, which by definition have activity that is primarily sublimation driven, and other active asteroids which do not \citep[e.g.,][]{2015Icar..248..289H,2017Natur.549..357A,2022arXiv220301397J}.

Other mechanisms for comet-like activity on small bodies include impact, as in the case of the NASA Double-Asteroid Redirection Test (DART) target (65803) Didymos-Dimorphos \citep{DART-AA} and main-belt asteroid (596) Scheila \citep{2012ApJ...744....9H}; rotational instability, as displayed by main-belt asteroids (6478) Gault \citep{chandlerSixYearsSustained2019} and (62412) 2000 SY$_{178}$ \citep{2015AJ....149...44S}; and thermal fracture, such as is hypothesized for near-Earth asteroids (3200) Phaethon and 2005 UD \citep{2013AJ....145..154L,2021Icar..36614535M}.

Thermal fracture is primarily applicable for near-Earth asteroids that experience large temperature gradients (several hundred degrees) over their orbits \citep[see][and references therein]{chandlerMigratoryOutburstingQuasiHilda2022,2022arXiv220301397J}. Additionally, observations of objects that are likely candidates for thermal fracture driven activity have found either a lack of evidence for activity for 2005 UD \citep{2023PSJ.....4...56K}, or, for (3200) Phaethon, that the observed activity is likely associated with gas emission rather than thermal fracture \citep{2023AJ....165...94H}; hence, this is an unlikely mechanism to explain the activity seen on \objname{}.

While both impact and rotational instability are difficult to rule out as drivers for activity (especially because the rotational period is unknown), neither of these mechanisms are directly correlated with perihelion. Hence, due to our discovery of activity on \objname{} at or near two separate perihelion passages, we conclude that sublimation of volatiles is the most likely cause for the observed activity on \objname{} \citep[see, for example,][]{2012ApJ...744....9H}.

\section{Summary and Future Work} \label{sec:sum}
Through our NASA Partner Citizen Science project {\it Active Asteroids} \citep[described in][]{chandlerChasingTailsActive2022}, we have discovered cometary activity emanating from quasi-Hilda \objname{} \citep{2023RNAAS...7...42O}. This activity occurred near the perihelion passage of \objname{} in 2016. Following this discovery, we conducted follow-up observations of \objname{} using the 3.5 m ARC telescope at Apache Point Observatory, Sunspot, New Mexico, USA and the 6.5 m Magellan Baade telescope at Las Campanas Observatory, Chile. From these observations, we discovered a second epoch of activity on \objname{}. This new epoch of activity occurred during the 2023 perihelion passage of \objname{}, approximately one orbital period after the first epoch detected. We performed a photometric analysis of the tail and find that it had similar apparent lengths and surfaces brightnesses in both epochs. Representative images from both epochs of activity are shown in Figure \ref{fig:images}.

We conducted dynamical simulations of \objname{} using $N$-body integration of orbital clones to determine probable orbital outcomes. Our simulations show that \objname{} experiences frequent close encounters with Jupiter over $\pm$1,000 yr (Figure \ref{fig:dynamics} (b)). 
These encounters perturb the orbit of \objname{} causing slight changes in its Tisserand parameter with respect to Jupiter allowing it to cross the traditional asteroid-comet boundary of $T_\mathrm{J}=3$ dozens of times on this timescale.
This causes a largely-superficial change in the orbital class of \objname{} over a 1,000 yr time period, with the potential for more substantial orbital migration over 10 kyr. During this time, JFC orbits are the most common over $\pm10$ kyr, with Centaur orbits being nearly as probably (Table \ref{tab:outcomes}).

Currently, \objname{} sits slightly outside of the quasi-Hilda semi-major axis range given in \citet{2006A&A...448.1191T}. However, because it is dynamically similar to other known quasi-Hildas, we classify \objname{} as a quasi-Hilda (Figure \ref{fig:corotation}).

We find the most probable cause for the activity on \objname{} is sublimation of volatile ices. While other mechanisms, such as rotational instability, could potentially cause the observed activity they are not correlated with perihelion. Therefore, since both epochs of detected activity are closely associated with perihelion passages, sublimation is the most likely cause.

Further observations of \objname{} will be particularly useful for characterizing the tail; for example, obtaining colors, monitoring its photometric evolution, and measuring surface brightness profiles. The remainder of this observing window, until $\sim$mid October 2023, is an ideal time for studying activity on \objname{}, since it is near perihelion. Additionally, future observations in coming years when \objname{} is not expected to be active will also be useful for comparative studies of the tail and nucleus, as well as for obtaining colors of the nucleus and measuring its rotational period. \objname{} will next reach perihelion in January 2030. We predict that \objname{}, after a period of inactivity following the recent perihelion passage, will reactivate as it approaches this date.

\section*{Acknowledgements}


We thank our anonymous reviewer who provided feedback that enhanced this work. We thank Hal Levison (SwRI) for helpful comments regarding dynamical classifications. We express our gratitude to Mark Magbanua (UCSF) for his frequent feedback to the {\it Active Asteroids} project. We also thank Elizabeth Baeten (Leuven, Belgium) for moderating the {\it Active Asteroids} forums and Cliff Johnson ({\it Zooniverse}) and Marc Kuchner (NASA) for their support, guidance, and feedback throughout this project. We thank Chris Coffey (NAU) and the NAU High Performance Computing Support team, who have made this work possible. We thank Jessica Birky (UW) and David Wang (UW) for contributing telescope time at APO.

We are grateful to the thousands of {\it Active Asteroids} volunteers who are critical in this process of discovery. We specifically thank those who helped to classify \objname{}: 
Al Lamperti (Royersford, USA),
Angelina A. Reese (Sequim, USA),
Dr.~Brian Leonard Goodwin (London, UK),
Clara Garza	(West Covina, USA),
C. M. Kaiser (Parker, USA),
Dawn Boles (Bakersfield, USA),
Eric Fabrigat (Velaux, France),
Ernest Jude P. Tiu	(Pototan, Philippines),
Frederick Hopper (Cotgrave, UK),
Henryk Krawczyk (Czeladż, Poland),
Ivan A. Terentev (Petrozavodsk, Russia),
Ivan Vladimirovich Sergienko (Sergiyev Posad, Russia),
J. Williams	(Swainsboro, USA),
Jayanta Ghosh (Purulia, India),
Jose A. da Silva Campos (Portugal),
Konstantinos Dimitrios Danalis	(Athens, Greece),
Martin Welham (Yatton, UK),
Marvin W. Huddleston (Mesquite, USA),
Michele T. Mazzucato (Florence, Italy),
Milton K. D. Bosch MD (Napa, USA),
Robert Pickard (Grove Hill, USA),
Robert Zach Moseley	(Worcester, USA),
Sarah Grissett (Tallahassee, USA),
Shelley-Anne Lake (Johannesburg, South Africa),
Somsikova Liudmila Leonidovna (Chirchik, Uzbekistan),
Stikhina Olga Sergeevna (Tyumen, Russia),
Thorsten Eschweiler (Übach-Palenberg, Germany),
Tiffany Shaw-Diaz (Dayton, USA),
Virgilio Gonano (Udine, Italy),
and
@WRSunset (Shaftesbury, UK).
We also thank super-classifier
C.J.A. Dukes (Oxford, UK).

This work was funded in part by NASA grant 80NSSC21K0114 (W.J.O. \& C.A.T.). C.O.C., H.H.H., and C.A.T.\ acknowledge support from the NASA Solar System Observations program (grant 80NSSC19K0869). This material is based upon work supported by the NSF Graduate Research Fellowship Program under grant No.\ 2018258765 and grant No.\ 2020303693. This work was supported in part by NSF award 1950901 (NAU REU program in astronomy and planetary science). Any opinions, findings, and conclusions or recommendations expressed in this material are those of the author(s) and do not necessarily reflect the views of the NSF. This work was made possible in part through the State of Arizona Technology \& Research Initiative Program.

Computational analyses were run on Northern Arizona University’s Monsoon computing cluster, funded by Arizona’s Technology and Research Initiative Fund. This research has made use of NASA’s Astrophysics Data System. This research has made use of data and/or services provided by the International Astronomical Union's Minor Planet Center. This research has made use of data and services provided by JPL Horizons \citep{giorginiJPLOnLineSolar1996}. This work made use of AstOrb, the Lowell Observatory Asteroid Orbit Database \textit{astorbDB} \citep{bowellPublicDomainAsteroid1994,moskovitzAstorbDatabaseLowell2021}. World Coordinate System corrections were facilitated by \textit{Astrometry.net} \citep{langAstrometryNetBlind2010}. This research has made use of The Institut de M\'ecanique C\'eleste et de Calcul des \'Eph\'em\'erides SkyBoT Virtual Observatory tool \citep{berthierSkyBoTNewVO2006}. This research is based on data obtained from the Astro Data Archive at NSF’s NOIRLab. These data are associated with observing programs 2016A-0189 (PI A. Rest) and 2015A-0121 (PI A. von der Linden). NOIRLab is managed by the Association of Universities for Research in Astronomy (AURA) under a cooperative agreement with the National Science Foundation. Based on observations obtained with the Apache Point Observatory 3.5-meter telescope, which is owned and operated by the Astrophysical Research Consortium. This paper includes data gathered with the 6.5 meter Magellan Telescopes located at Las Campanas Observatory, Chile.

\vspace{8pt}
\facilities{
APO: 3.5m (ARCTIC), 
CTIO: 4m Blanco (DECam),
LCO: 6.5m Magellan Baade (IMACS).\\
}

\software{
        {\tt acronym} \citep{2017JOSS....2..102L},
        {\tt astropy} \citep{robitailleAstropyCommunityPython2013}, 
        {\tt Matplotlib} \citep{hunterMatplotlib2DGraphics2007},
        {\tt NumPy} \citep{harrisArrayProgrammingNumPy2020},
        {\tt pandas} \citep{rebackPandasdevPandasPandas2022},
        {\tt Photutils} \citep{larry_bradley_2023_7946442},
        {\tt REBOUND} \citep{2012A&A...537A.128R,2015MNRAS.446.1424R},
        {\tt SAOImageDS9} \citep{joyeNewFeaturesSAOImage2006},
        {\tt SciPy} \citep{virtanenSciPyFundamentalAlgorithms2020},
        \texttt{termcolor} (\url{https://pypi.org/project/termcolor/}),
        \texttt{tqdm} \citep{casper_da_costa_luis_2020_4293724},
        \texttt{VizieR} \citep{ochsenbeinVizieRDatabaseAstronomical2000}.
          }

\bibliography{arXivSubmission}{}

\begin{thebibliography}{}
\expandafter\ifx\csname natexlab\endcsname\relax\def\natexlab#1{#1}\fi
\providecommand{\url}[1]{\href{#1}{#1}}
\providecommand{\dodoi}[1]{doi:~\href{http://doi.org/#1}{\nolinkurl{#1}}}
\providecommand{\doeprint}[1]{\href{http://ascl.net/#1}{\nolinkurl{http://ascl.net/#1}}}
\providecommand{\doarXiv}[1]{\href{https://arxiv.org/abs/#1}{\nolinkurl{https://arxiv.org/abs/#1}}}

\bibitem[{{Agarwal} {et~al.}(2017){Agarwal}, {Jewitt}, {Mutchler}, {Weaver}, \& {Larson}}]{2017Natur.549..357A}
{Agarwal}, J., {Jewitt}, D., {Mutchler}, M., {Weaver}, H., \& {Larson}, S. 2017, \nat, 549, 357, \dodoi{10.1038/nature23892}

\bibitem[{Berthier {et~al.}(2006)Berthier, Vachier, Thuillot, Fernique, Ochsenbein, Genova, Lainey, Arlot, Gabriel, Arviset, Ponz, \& Solano}]{berthierSkyBoTNewVO2006}
Berthier, J., Vachier, F., Thuillot, W., {et~al.} 2006, in Astronomical {{Data Analysis Software}} and {{Systems XV ASP Conference Series}}, Vol. 351 ({Orem, UT}: {Astronomical Society of the Pacific}), 367

\bibitem[{Bowell {et~al.}(1994)Bowell, Muinonen, \& Wasserman}]{bowellPublicDomainAsteroid1994}
Bowell, E., Muinonen, K., \& Wasserman, L.~H. 1994, Symposium-International \ldots

\bibitem[{Bradley {et~al.}(2023)Bradley, Sip{\H o}cz, Robitaille, Tollerud, Vin{\'{\i}}cius, Deil, Barbary, Wilson, Busko, Donath, G{\"u}nther, Cara, Lim, Me{\ss}linger, Conseil, Bostroem, Droettboom, Bray, Bratholm, Barentsen, Craig, Rathi, Pascual, Perren, Georgiev, de~Val-Borro, Kerzendorf, Bach, Quint, \& Souchereau}]{larry_bradley_2023_7946442}
Bradley, L., Sip{\H o}cz, B., Robitaille, T., {et~al.} 2023, astropy/photutils: 1.8.0, 1.8.0,  Zenodo, \dodoi{10.5281/zenodo.7946442}

\bibitem[{{Brown} \& {Rein}(2023)}]{2023MNRAS.tmp..703B}
{Brown}, G., \& {Rein}, H. 2023, \mnras, \dodoi{10.1093/mnras/stad719}

\bibitem[{Chandler(2022)}]{chandlerChasingTailsActive2022}
Chandler, C.~O. 2022, PhD thesis, Northern Arizona University, {Flagstaff, Arizona, USA}

\bibitem[{Chandler {et~al.}(2018)Chandler, Curtis, Mommert, Sheppard, \& Trujillo}]{chandlerSAFARISearchingAsteroids2018}
Chandler, C.~O., Curtis, A.~M., Mommert, M., Sheppard, S.~S., \& Trujillo, C.~A. 2018, Publications of the Astronomical Society of the Pacific, 130, 114502, \dodoi{10.1088/1538-3873/aad03d}

\bibitem[{Chandler {et~al.}(2019)Chandler, Kueny, Gustafsson, Trujillo, Robinson, \& Trilling}]{chandlerSixYearsSustained2019}
Chandler, C.~O., Kueny, J., Gustafsson, A., {et~al.} 2019, The Astrophysical Journal Letters, 877, L12, \dodoi{10/gg3qw6}

\bibitem[{Chandler {et~al.}(2022)Chandler, Oldroyd, \& Trujillo}]{chandlerMigratoryOutburstingQuasiHilda2022}
Chandler, C.~O., Oldroyd, W.~J., \& Trujillo, C.~A. 2022, The Astrophysical Journal, 937, L2, \dodoi{10.3847/2041-8213/ac897a}

\bibitem[{{Chandler} {et~al.}(2023){Chandler}, {Oldroyd}, {Trujillo}, {Burris}, {Hsieh}, {Kueny}, {Mazzucato}, {Bosch}, \& {Shaw-Diaz}}]{2023RNAAS...7...27C}
{Chandler}, C.~O., {Oldroyd}, W.~J., {Trujillo}, C.~A., {et~al.} 2023, Research Notes of the American Astronomical Society, 7, 27, \dodoi{10.3847/2515-5172/acbbce}

\bibitem[{da~Costa-Luis {et~al.}(2020)da~Costa-Luis, Larroque, Altendorf, Mary, Korobov, Yorav-Raphael, Ivanov, Bargull, Rodrigues, CHEN, Newey, James, Zugnoni, Pagel, mjstevens777, Dektyarev, Rothberg, Alexander, Panteleit, Dill, FichteFoll, HeoHeo, van Kemenade, McCracken, Nordlund, Nechaev, Desh, RedBug312, richardsheridan, \& Socialery}]{casper_da_costa_luis_2020_4293724}
da~Costa-Luis, C., Larroque, S.~K., Altendorf, K., {et~al.} 2020, {tqdm: A fast, Extensible Progress Bar for Python and CLI}, v4.54.0,  Zenodo, \dodoi{10.5281/zenodo.4293724}

\bibitem[{{DePoy} {et~al.}(2008){DePoy}, {Abbott}, {Annis}, {Antonik}, {Barcel{\'o}}, {Bernstein}, {Bigelow}, {Brooks}, {Buckley-Geer}, {Campa}, {Cardiel}, {Castander}, {Castilla}, {Cease}, {Chappa}, {Dede}, {Derylo}, {Diehl}, {Doel}, {DeVicente}, {Estrada}, {Finley}, {Flaugher}, {Gaztanaga}, {Gerdes}, {Gladders}, {Guarino}, {Gutierrez}, {Hamilton}, {Haney}, {Holland}, {Honscheid}, {Huffman}, {Karliner}, {Kau}, {Kent}, {Kozlovsky}, {Kubik}, {Kuehn}, {Kuhlmann}, {Kuk}, {Leger}, {Lin}, {Martinez}, {Martinez}, {Merritt}, {Mohr}, {Moore}, {Moore}, {Nord}, {Ogando}, {Olsen}, {Onal}, {Peoples}, {Qian}, {Roe}, {Sanchez}, {Scarpine}, {Schmidt}, {Schmitt}, {Schubnell}, {Schultz}, {Selen}, {Shaw}, {Simaitis}, {Slaughter}, {Smith}, {Spinka}, {Stefanik}, {Stuermer}, {Talaga}, {Tarle}, {Thaler}, {Tucker}, {Walker}, {Worswick}, \& {Zhao}}]{2008SPIE.7014E..0ED}
{DePoy}, D.~L., {Abbott}, T., {Annis}, J., {et~al.} 2008, in Society of Photo-Optical Instrumentation Engineers (SPIE) Conference Series, Vol. 7014, Ground-based and Airborne Instrumentation for Astronomy II, ed. I.~S. {McLean} \& M.~M. {Casali}, 70140E, \dodoi{10.1117/12.789466}

\bibitem[{{Dressler} {et~al.}(2011){Dressler}, {Bigelow}, {Hare}, {Sutin}, {Thompson}, {Burley}, {Epps}, {Oemler}, {Bagish}, {Birk}, {Clardy}, {Gunnels}, {Kelson}, {Shectman}, \& {Osip}}]{2011PASP..123..288D}
{Dressler}, A., {Bigelow}, B., {Hare}, T., {et~al.} 2011, \pasp, 123, 288, \dodoi{10.1086/658908}

\bibitem[{{Garc{\'\i}a-Migani} \& {Gil-Hutton}(2018)}]{2018P&SS..160...12G}
{Garc{\'\i}a-Migani}, E., \& {Gil-Hutton}, R. 2018, \planss, 160, 12, \dodoi{10.1016/j.pss.2018.03.011}

\bibitem[{{Gil-Hutton} \& {Garc{\'\i}a-Migani}(2016)}]{2016A&A...590A.111G}
{Gil-Hutton}, R., \& {Garc{\'\i}a-Migani}, E. 2016, \aap, 590, A111, \dodoi{10.1051/0004-6361/201628184}

\bibitem[{Giorgini {et~al.}(1996)Giorgini, Yeomans, Chamberlin, Chodas, Jacobson, Keesey, Lieske, Ostro, Standish, \& Wimberly}]{giorginiJPLOnLineSolar1996}
Giorgini, J.~D., Yeomans, D.~K., Chamberlin, A.~B., {et~al.} 1996, American Astronomical Society, 28, 25.04

\bibitem[{Harris {et~al.}(2020)Harris, Millman, {van der Walt}, Gommers, Virtanen, Cournapeau, Wieser, Taylor, Berg, Smith, Kern, Picus, Hoyer, {van Kerkwijk}, Brett, Haldane, {del R{\'i}o}, Wiebe, Peterson, {G{\'e}rard-Marchant}, Sheppard, Reddy, Weckesser, Abbasi, Gohlke, \& Oliphant}]{harrisArrayProgrammingNumPy2020}
Harris, C.~R., Millman, K.~J., {van der Walt}, S.~J., {et~al.} 2020, Nature, 585, 357, \dodoi{10.1038/s41586-020-2649-2}

\bibitem[{{Hernandez} {et~al.}(2022){Hernandez}, {Zeebe}, \& {Hadden}}]{2022MNRAS.510.4302H}
{Hernandez}, D.~M., {Zeebe}, R.~E., \& {Hadden}, S. 2022, \mnras, 510, 4302, \dodoi{10.1093/mnras/stab3664}

\bibitem[{Hill(1878)}]{hillResearchesLunarTheory1878}
Hill, G.~W. 1878, American Journal of Mathematics, 1, 5, \dodoi{10.2307/2369430}

\bibitem[{{Hsieh} {et~al.}(2012){Hsieh}, {Yang}, \& {Haghighipour}}]{2012ApJ...744....9H}
{Hsieh}, H.~H., {Yang}, B., \& {Haghighipour}, N. 2012, \apj, 744, 9, \dodoi{10.1088/0004-637X/744/1/9}

\bibitem[{{Hsieh} {et~al.}(2015){Hsieh}, {Denneau}, {Wainscoat}, {Sch{\"o}rghofer}, {Bolin}, {Fitzsimmons}, {Jedicke}, {Kleyna}, {Micheli}, {Vere{\v{s}}}, {Kaiser}, {Chambers}, {Burgett}, {Flewelling}, {Hodapp}, {Magnier}, {Morgan}, {Price}, {Tonry}, \& {Waters}}]{2015Icar..248..289H}
{Hsieh}, H.~H., {Denneau}, L., {Wainscoat}, R.~J., {et~al.} 2015, \icarus, 248, 289, \dodoi{10.1016/j.icarus.2014.10.031}

\bibitem[{{Huehnerhoff} {et~al.}(2016){Huehnerhoff}, {Ketzeback}, {Bradley}, {Dembicky}, {Doughty}, {Hawley}, {Johnson}, {Klaene}, {Leon}, {McMillan}, {Owen}, {Sayres}, {Sheen}, \& {Shugart}}]{2016SPIE.9908E..5HH}
{Huehnerhoff}, J., {Ketzeback}, W., {Bradley}, A., {et~al.} 2016, in Society of Photo-Optical Instrumentation Engineers (SPIE) Conference Series, Vol. 9908, Ground-based and Airborne Instrumentation for Astronomy VI, ed. C.~J. {Evans}, L.~{Simard}, \& H.~{Takami}, 99085H, \dodoi{10.1117/12.2234214}

\bibitem[{{Hui}(2023)}]{2023AJ....165...94H}
{Hui}, M.-T. 2023, \aj, 165, 94, \dodoi{10.3847/1538-3881/acae9c}

\bibitem[{Hunter(2007)}]{hunterMatplotlib2DGraphics2007}
Hunter, J.~D. 2007, Computing in Science \& Engineering, 9, 90, \dodoi{10.1109/MCSE.2007.55}

\bibitem[{Jewitt {et~al.}(2015)Jewitt, Hsieh, \& Agarwal}]{jewittActiveAsteroids2015}
Jewitt, D., Hsieh, H., \& Agarwal, J. 2015, in Asteroids {{IV}} ({Tucson, Arizona}: {University of Arizona Press}), 221--241

\bibitem[{{Jewitt} \& {Hsieh}(2022)}]{2022arXiv220301397J}
{Jewitt}, D., \& {Hsieh}, H.~H. 2022, arXiv e-prints, arXiv:2203.01397, \dodoi{10.48550/arXiv.2203.01397}

\bibitem[{{Jewitt} \& {Kim}(2020)}]{2020PSJ.....1...77J}
{Jewitt}, D., \& {Kim}, Y. 2020, \psj, 1, 77, \dodoi{10.3847/PSJ/abbef6}

\bibitem[{{Jewitt} {et~al.}(2019){Jewitt}, {Kim}, {Rajagopal}, {Ridgway}, {Kotulla}, {Liu}, {Mutchler}, {Li}, {Weaver}, \& {Larson}}]{2019AJ....157...54J}
{Jewitt}, D., {Kim}, Y., {Rajagopal}, J., {et~al.} 2019, \aj, 157, 54, \dodoi{10.3847/1538-3881/aaf563}

\bibitem[{{Jordi} {et~al.}(2006){Jordi}, {Grebel}, \& {Ammon}}]{2006A&A...460..339J}
{Jordi}, K., {Grebel}, E.~K., \& {Ammon}, K. 2006, \aap, 460, 339, \dodoi{10.1051/0004-6361:20066082}

\bibitem[{Joye(2006)}]{joyeNewFeaturesSAOImage2006}
Joye, W.~A. 2006, in Astronomical {{Data Analysis Software}} and {{Systems XV ASP Conference Series}}, Vol. 351, 574--

\bibitem[{{Kres{\'a}k}(1972)}]{1972IAUS...45..503K}
{Kres{\'a}k}, L. 1972, in The Motion, Evolution of Orbits, and Origin of Comets, ed. G.~A. {Chebotarev}, E.~I. {Kazimirchak-Polonskaia}, \& B.~G. {Marsden}, Vol.~45, 503

\bibitem[{{Kueny} {et~al.}(2023){Kueny}, {Chandler}, {Devog{\'e}le}, {Moskovitz}, {Pravec}, {Ku{\v{c}}{\'a}kov{\'a}}, {Hornoch}, {Ku{\v{s}}nir{\'a}k}, {Granvik}, {Konstantopoulou}, {Jannsen}, {Moran}, {Siltala}, {Fedorets}, {Ferrais}, {Jehin}, {Kareta}, \& {Hanu{\v{s}}}}]{2023PSJ.....4...56K}
{Kueny}, J.~K., {Chandler}, C.~O., {Devog{\'e}le}, M., {et~al.} 2023, \psj, 4, 56, \dodoi{10.3847/PSJ/acc1e7}

\bibitem[{Lang {et~al.}(2010)Lang, Hogg, Mierle, Blanton, \& Roweis}]{langAstrometryNetBlind2010}
Lang, D., Hogg, D.~W., Mierle, K., Blanton, M., \& Roweis, S. 2010, Astronomical Journal, 139, 1782, \dodoi{10.1088/0004-6256/139/5/1782}

\bibitem[{{Levison}(1996)}]{1996ASPC..107..173L}
{Levison}, H.~F. 1996, in Astronomical Society of the Pacific Conference Series, Vol. 107, Completing the Inventory of the Solar System, ed. T.~{Rettig} \& J.~M. {Hahn}, 173--191

\bibitem[{{Li} \& {Jewitt}(2013)}]{2013AJ....145..154L}
{Li}, J., \& {Jewitt}, D. 2013, \aj, 145, 154, \dodoi{10.1088/0004-6256/145/6/154}

\bibitem[{{Li} {et~al.}(2023){Li}, {Hirabayashi}, {Farnham}, {Sunshine}, {Knight}, {Tancredi}, {Moreno}, {Murphy}, {Opitom}, {Chesley}, {Scheeres}, {Thomas}, {Fahnestock}, {Cheng}, {Dressel}, {Ernst}, {Ferrari}, {Fitzsimmons}, {Ieva}, {Ivanovski}, {Kareta}, {Kolokolova}, {Lister}, {Raducan}, {Rivkin}, {Rossi}, {Soldini}, {Stickle}, {Vick}, {Vincent}, {Weaver}, {Bagnulo}, {Bannister}, {Cambioni}, {Campo Bagatin}, {Chabot}, {Cremonese}, {Terik Daly}, {Dotto}, {Glenar}, {Granvik}, {Hasselmann}, {Herreros}, {Jacobson}, {Jutzi}, {Kohout}, {La Forgia}, {Lazzarin}, {Lin}, {Lolachi}, {Lucchetti}, {Makadia}, {Mazzotta Epifani}, {Michel}, {Migliorini}, {Moskovitz}, {Orm.}, {Pajola}, {nchez}, {Schwartz}, {Snodgrass}, {Steckloff}, {Stubbs}, \& {Trigo-Rodriguez}}]{DART-AA}
{Li}, J.-Y., {Hirabayashi}, M., {Farnham}, T.~L., {et~al.} 2023, Nature, 616, 452, \dodoi{10.1038/s41586-023-05811-4}

\bibitem[{{MacLennan} {et~al.}(2021){MacLennan}, {Toliou}, \& {Granvik}}]{2021Icar..36614535M}
{MacLennan}, E., {Toliou}, A., \& {Granvik}, M. 2021, \icarus, 366, 114535, \dodoi{10.1016/j.icarus.2021.114535}

\bibitem[{{Morbidelli} {et~al.}(2000){Morbidelli}, {Chambers}, {Lunine}, {Petit}, {Robert}, {Valsecchi}, \& {Cyr}}]{2000M&PS...35.1309M}
{Morbidelli}, A., {Chambers}, J., {Lunine}, J.~I., {et~al.} 2000, \maps, 35, 1309, \dodoi{10.1111/j.1945-5100.2000.tb01518.x}

\bibitem[{Moskovitz {et~al.}(2021)Moskovitz, Burt, Schottland, Wasserman, Bailen, Grimm, \& Granvik}]{moskovitzAstorbDatabaseLowell2021}
Moskovitz, N., Burt, B., Schottland, R., {et~al.} 2021, AAS Division of Planetary Science meeting \#53, id. 101.04, 53, 101.04

\bibitem[{{O'Brien} {et~al.}(2018){O'Brien}, {Izidoro}, {Jacobson}, {Raymond}, \& {Rubie}}]{2018SSRv..214...47O}
{O'Brien}, D.~P., {Izidoro}, A., {Jacobson}, S.~A., {Raymond}, S.~N., \& {Rubie}, D.~C. 2018, \ssr, 214, 47, \dodoi{10.1007/s11214-018-0475-8}

\bibitem[{Ochsenbein {et~al.}(2000)Ochsenbein, Bauer, \& Marcout}]{ochsenbeinVizieRDatabaseAstronomical2000}
Ochsenbein, F., Bauer, P., \& Marcout, J. 2000, Astronomy and Astrophysics Supplement, 143, 23, \dodoi{10/fb95hg}

\bibitem[{{Ohtsuka} {et~al.}(2008){Ohtsuka}, {Ito}, {Yoshikawa}, {Asher}, \& {Arakida}}]{2008A&A...489.1355O}
{Ohtsuka}, K., {Ito}, T., {Yoshikawa}, M., {Asher}, D.~J., \& {Arakida}, H. 2008, \aap, 489, 1355, \dodoi{10.1051/0004-6361:200810321}

\bibitem[{{Oldroyd} {et~al.}(2023){Oldroyd}, {Chandler}, {Trujillo}, {Burris}, {Kueny}, {Hsieh}, {Farrell}, {DeSpain}, {Mazzucato}, {Bosch}, {Shaw-Diaz}, \& {Gonano}}]{2023RNAAS...7...42O}
{Oldroyd}, W.~J., {Chandler}, C.~O., {Trujillo}, C.~A., {et~al.} 2023, Research Notes of the American Astronomical Society, 7, 42, \dodoi{10.3847/2515-5172/acc17c}

\bibitem[{{Pravec} {et~al.}(2022){Pravec}, {Thomas}, {Rivkin}, {Scheirich}, {Moskovitz}, {Knight}, {Snodgrass}, {de Le{\'o}n}, {Licandro}, {Popescu}, {Thirouin}, {F{\"o}hring}, {Chandler}, {Oldroyd}, {Trujillo}, {Howell}, {Green}, {Thomas-Osip}, {Sheppard}, {Farnham}, {Mazzotta Epifani}, {Dotto}, {Ieva}, {Dall'Ora}, {Kokotanekova}, {Carry}, \& {Souami}}]{2022PSJ.....3..175P}
{Pravec}, P., {Thomas}, C.~A., {Rivkin}, A.~S., {et~al.} 2022, \psj, 3, 175, \dodoi{10.3847/PSJ/ac7be1}

\bibitem[{Reback {et~al.}(2022)Reback, {jbrockmendel}, McKinney, den Bossche, Augspurger, Roeschke, Hawkins, Cloud, {gfyoung}, Sinhrks, Hoefler, Klein, Petersen, Tratner, She, Ayd, Naveh, Darbyshire, Garcia, Shadrach, Schendel, Hayden, Saxton, Gorelli, Li, Zeitlin, Jancauskas, McMaster, W{\"o}rtwein, \& Battiston}]{rebackPandasdevPandasPandas2022}
Reback, J., {jbrockmendel}, McKinney, W., {et~al.} 2022, Pandas-Dev/Pandas: {{Pandas}} 1.4.2, Zenodo, \dodoi{10.5281/zenodo.6408044}

\bibitem[{{Rein} \& {Liu}(2012)}]{2012A&A...537A.128R}
{Rein}, H., \& {Liu}, S.~F. 2012, \aap, 537, A128, \dodoi{10.1051/0004-6361/201118085}

\bibitem[{{Rein} \& {Spiegel}(2015)}]{2015MNRAS.446.1424R}
{Rein}, H., \& {Spiegel}, D.~S. 2015, \mnras, 446, 1424, \dodoi{10.1093/mnras/stu2164}

\bibitem[{Robitaille {et~al.}(2013)Robitaille, Tollerud, Greenfield, Droettboom, Bray, Aldcroft, Davis, Ginsburg, {Price-Whelan}, Kerzendorf, Conley, Crighton, Barbary, Muna, Ferguson, Grollier, Parikh, Nair, G{\"u}nther, Deil, Woillez, Conseil, Kramer, Turner, Singer, Fox, Weaver, Zabalza, Edwards, Azalee~Bostroem, Burke, Casey, Crawford, Dencheva, Ely, Jenness, Labrie, Lim, Pierfederici, Pontzen, Ptak, Refsdal, Servillat, \& Streicher}]{robitailleAstropyCommunityPython2013}
Robitaille, T.~P., Tollerud, E.~J., Greenfield, P., {et~al.} 2013, Astronomy \& Astrophysics, 558, A33, \dodoi{10/gfvntd}

\bibitem[{{Sheppard} \& {Trujillo}(2015)}]{2015AJ....149...44S}
{Sheppard}, S.~S., \& {Trujillo}, C. 2015, \aj, 149, 44, \dodoi{10.1088/0004-6256/149/2/44}

\bibitem[{{Tiscareno} \& {Malhotra}(2003)}]{2003AJ....126.3122T}
{Tiscareno}, M.~S., \& {Malhotra}, R. 2003, \aj, 126, 3122, \dodoi{10.1086/379554}

\bibitem[{{Toth}(2006)}]{2006A&A...448.1191T}
{Toth}, I. 2006, \aap, 448, 1191, \dodoi{10.1051/0004-6361:20053492}

\bibitem[{{Trujillo} {et~al.}(2023){Trujillo}, {Chandler}, {Oldroyd}, {Burris}, {Hsieh}, {Kueny}, {Mazzucato}, {Bosch}, {Shaw-Diaz}, \& {Gonano}}]{2023RNAAS...7..106T}
{Trujillo}, C.~A., {Chandler}, C.~O., {Oldroyd}, W.~J., {et~al.} 2023, Research Notes of the American Astronomical Society, 7, 106, \dodoi{10.3847/2515-5172/acd7f0}

\bibitem[{Virtanen {et~al.}(2020)Virtanen, Gommers, Oliphant, Haberland, Reddy, Cournapeau, Burovski, Peterson, Weckesser, Bright, {van der Walt}, Brett, Wilson, Millman, Mayorov, Nelson, Jones, Kern, Larson, Carey, Polat, Feng, Moore, VanderPlas, Laxalde, Perktold, Cimrman, Henriksen, Quintero, Harris, Archibald, Ribeiro, Pedregosa, \& {van Mulbregt}}]{virtanenSciPyFundamentalAlgorithms2020}
Virtanen, P., Gommers, R., Oliphant, T.~E., {et~al.} 2020, Nature Methods, 17, 261, \dodoi{10.1038/s41592-019-0686-2}

\bibitem[{{Weisenburger} {et~al.}(2017){Weisenburger}, {Huehnerhoff}, {Levesque}, \& {Massey}}]{2017JOSS....2..102L}
{Weisenburger}, K., {Huehnerhoff}, J., {Levesque}, E., \& {Massey}, P. 2017, The Journal of Open Source Software, 2, 102, \dodoi{10.21105/joss.00102}

\bibitem[{{Wisdom}(2015)}]{2015AJ....150..127W}
{Wisdom}, J. 2015, \aj, 150, 127, \dodoi{10.1088/0004-6256/150/4/127}

\end{thebibliography}
\bibliographystyle{aasjournal}

\end{document}